# Coexistence and interplay of two ferroelectric mechanisms in Zn$_{1-x}$Mg$_x$O

Jonghee Yang,[1,2] Anton V. Ievlev,[3] Anna N. Morozovska,[4] Eugene Eliseev,[4] Jonathan D Poplawsky,[3] Devin Goodling,[5] Robert Jackson Spurling,[5] Jon-Paul Maria,[5] Sergei V. Kalinin,[1*] Yongtao Liu[3*]

[1] Department of Materials Science and Engineering, University of Tennessee, Knoxville, TN 37996, USA

[2] Department of Chemistry, Yonsei University, Seoul 03722, Republic of Korea

[3] Center for Nanophase Materials Sciences, Oak Ridge National Laboratory, Oak Ridge, TN 37830, USA

[4] Institute of Physics, National Academy of Sciences of Ukraine, 46, pr. Nauky, 03028 Kyiv, Ukraine

[5] Materials Research Institute, Pennsylvania State University, University Park, PA 16802, USA

Corresponding emails: sergei2@utk.edu; liuy3@ornl.gov

**Abstract**

Ferroelectric materials promise exceptional attributes including low power dissipation, fast operational speeds, enhanced endurance, and superior retention to revolutionize information technology. However, the practical application of ferroelectric-semiconductor memory devices has been significantly challenged by the incompatibility of traditional perovskite oxide ferroelectrics with metal-oxide-semiconductor technology. Recent discoveries of ferroelectricity in binary oxides such as $Zn_{1-x}Mg_xO$ and $Hf_{1-x}Zr_xO$ have been a focal point of research in ferroelectric information technology. This work investigates the ferroelectric properties of $Zn_{1-x}Mg_xO$ utilizing automated band excitation piezoresponse force microscopy. Our findings reveal the coexistence of two ferroelectric subsystems within $Zn_{1-x}Mg_xO$. We propose a "fringing-ridge mechanism" of polarization switching that is characterized by initial lateral expansion of nucleation without significant propagation in depth, contradicting the conventional domain growth process observed in ferroelectrics. This unique polarization dynamics in $Zn_{1-x}Mg_xO$ suggests a new understanding of ferroelectric behavior, contributing to both the fundamental science of ferroelectrics and their application in information technology.

Ferroelectric materials have been a focal point of research for over five decades because of their potential applications in information technology.[1, 2] Ferroelectrics with two or more switchable polarization states separated by atomically thin walls are promising for high-density information storage, e.g., ferroelectric random-access memories (FeRAM)[3, 4] or tunneling barriers. These in turn are believed to offer a constellation of exceptional attributes encompassing low power dissipation, fast operational speeds, enhanced endurance, and good retention. However, the practical implementation of ferroelectric-semiconductor memory has been hindered by the fact that metal-oxide-semiconductor (CMOS) technology is generally incompatible with the film growth of traditional perovskite oxide ferroelectrics. Despite over three decades of research towards the integration of classical ferroelectrics and semiconductors, sustained and scalable solutions have not yet been found.

The discovery of ferroelectricity within hafnia solid solutions has been a disruptive change.[5] In contrast to their perovskite counterparts, these materials can be processed via deposition techniques such as atomic layer deposition, which enables seamless integration into established semiconductor fabrication workflows.[6, 7] As a consequence, the last decade has witnessed extensive research endeavors in the exploration of ferroelectric properties across a wide range of materials.[7] Beyond hafnia oxide, ferroelectricity has also been unveiled in wurtzite such as $Al_xSc_{1-x}N$,[8] $Al_xB_{1-x}N$,[9-11] and $Zn_{1-x}Mg_xO$.[12] However, these emergent ferroelectric wurtzite materials exhibit distinctive behavior deviating from traditional perovskite ferroelectrics. These unique behaviors include their square ferroelectric hysteresis loops even in the presence of strong structural disorder, the 'wake-up' effect where the ferroelectric response emerges or enhances after several polling cycles,[9, 13-15] and the potential absence of a paraelectric phase. The 'wake-up' effect in $Zn_{1-x}Mg_xO$ has been found field-dependent, where $Zn_{1-x}Mg_xO$ can be woken up in 20 cycles on driving near the coercive field and in a single cycle on driving in excess of 4 MV/cm.[14] In addition, strong electrochemical reactions were observed when applying an electric field on the bare surface of $Al_xB_{1-x}N$ in ambient condition, such electrochemical reactions can be suppressed in a controlled environment with low humidity.[16] Recently, it was also discovered that the atmosphere affects the polarization stability in a binary oxide $Hf_{0.5}Zr_{0.5}O$ film.[17]

In this work, we investigate the polarization switching in $Zn_{1-x}Mg_xO$ with band excitation piezoresponse force microscopy (BEPFM) imaging and band excitation piezoresponse spectroscopy (BEPS) measurements in a high throughput manner to thoroughly visualize the polarization dynamics. We observe highly unusual ferroelectric behavior that suggests the coexistence of two independent ferroelectric subsystems, the behavior is inconsistent with strong depolarization field coupling. We explore the detailed mechanism of this phenomena and argue that we observe the evidence of a new type of ferroelectric switching—fringing-ridge mechanism correlated with incomplete switching. Unlike a traditional domain growth process, where the domain propagates to depth after nucleation, in a fringing-ridge mechanism, nucleation expands laterally first with rare or without propagation to depth; domain propagation only happens at very high voltage.

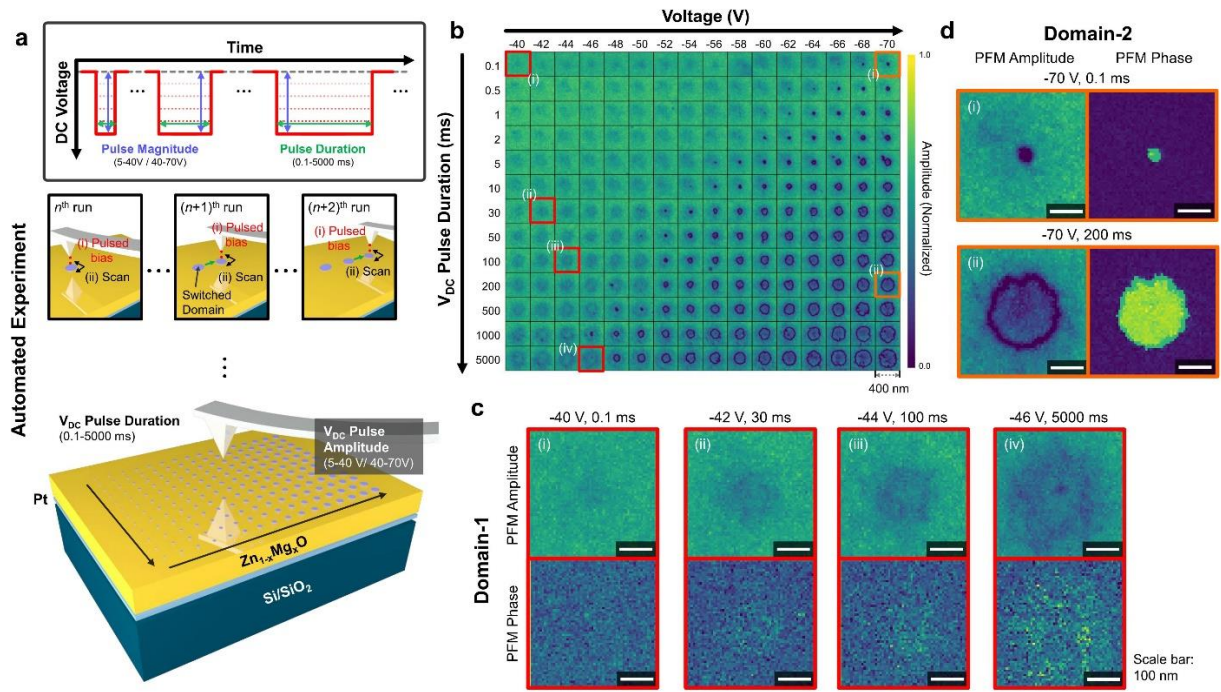

**Figure 1.** High-throughput domain writing experiments. (a) Scheme of experimental process, where a DC pulse is applied to reverse the polarization and subsequently a PFM image measurement reveals the resultant domain created by the DC pulse; varying the DC pulse parameters (i.e., duration and magnitude) allows to explore the domain growth procedure. (b) High-throughput domain writing results of ZMO-32, where 16 pulse magnitude and 13 pulse duration conditions are explored (208 pulse conditions). The results show two distinct domain structures including shadow domains and normal domains. (c), (d) show amplified images of representative domains.

To explore polarization switching in this material, we perform the exploration of the domain growth dynamics as a function of the bias pulse parameters for three $Zn_{1-x}Mg_xO$ samples with different compositions, targeting the exploration of the composition dependent domain growth laws. The film growth procedure can be found in Methods section. Hereinafter, we refer to these three samples as ZMO-32, ZMO-24, ZMO-18, respectively, according to the Mg composition in the films, i.e., Mg ratio in ZMO-32 is 32%. The as-grown ZMO films show a down-polarized pristine state. The polarization can be switched by applying a direct current (DC) bias via a scanning probe microscopy tip. BEPFM mappings of poled ZMO films displaying the standard ferroelectric domains are shown in Suppl. Mat. Figure S1.

To accelerate the exploration of the large parameter spaces corresponding to the possible switching conditions, we developed a high throughout domain writing and imaging approach based on using AEcroscoPy,[18, 19] a universal platform for automated experiments in scanning probe and electron microscopy. As shown in Figure 1a, this experiment starts with applying a DC pulse at the center of the image area to switch the polarization. As a next step, a high resolution BEPFM image measurement is performed to map the polarization state of this region. Then, the same measurement is performed at a different location by applying a DC pulse with a modified magnitude and/or duration to study the effect of DC pulse parameters on polarization switching. Whether the polarization can be switched depends on the magnitude and duration of the applied DC pulse, often larger and longer DC pulse can lead to polarization nucleation and reversal, and subsequent expansion of the polarization reversed domain, *i.e.*, domain growth. Here, we can modify the pulse parameters and perform BEPFM maps in an automated manner, allowing us to conduct high throughput experimentation to systematically study the domain growth dependence on pulse condition; also, a subsequent high resolution PFM image after each pulse allows visualization of more detailed local polarization state compared to an entire observing scan after applying all pulses.

We first study the ZMO-32 film by applying DC pulses with a magnitude ranging from negative 40 V-70 V and a duration ranging from 0.1 ms to 5000 ms. Figure 1b shows the BEPFM amplitude results (corresponding phase results can be found in Figure S2). Each square in Figure 1b represents an independent BEPFM measurement after applying different DC pulses, where the corresponding DC pulse magnitude and duration are labeled. Here 13 pulse durations and 16 pulse magnitudes, totaling 208 pulse conditions, are investigated. We note that previous similar experiments have been used for probing domain growth in materials such as LiNbO$_3$,[20-22] Hf$_{0.5}$Zr$_{0.5}$O,[23] and BaTiO$_3$[24], and can be used to construct domain size as a function of pulse parameters.

However, exploration of the domain switching in ZMO have led to the observation of very unusual domain dynamics. For this material, we observed two kinds of resulted domain structures, one is a shadow domain (domain-1) at the top left, and the other is a normal domain structure (domain-2) at the bottom right that shows dark amplitude contrast at domain walls. Several representative domain-1 amplitude and phase images are amplified in Figure 1c. Under low and/or short bias (e.g. Figure 1c-i), domain-1 shows weak changes in amplitude images appearing as dark shadow and without apparent domain walls, and no apparent changes in phase images. With bias increase (*e.g.* Figure 1c-iv), domain-1 can adopt fairly complex geometry with extremely rough domain walls. Increasing the pulse magnitude or duration will lead to the formation of domain-2 structures, consistent with normal polarization switching in uniaxial ferroelectric. Two representative domain-2 amplitude and phase images are shown in Figure 1d, domain-2 shows apparent domain walls and phase inversion. It seems that domain-2 overwrites domain-1, thus domain-1 becomes less visible when domain-2 forms. We performed another high throughput domain writing experiment with small DC pulses ranging from 5 V-40 V to specifically study evolution of shape and size of domain-1 structures. Results are shown in Figure S3, it is seen that the domain-1 can form when the bias is as low as 12.5 V, which is much lower than the known coercive voltage (~55 V for these samples) from macroscopic measurement of this sample. In addition, the size of domain-1 can reach hundreds of nanometers. Similar domain writing experiments are performed on ZMO-24 and ZMO-18 films, the results are shown in Figure S4-S7. Two domain states are highly reproduced in films with different composition.

The immediate interpretation of the shadow domains would be surface charging, giving rise to the domain-like contrast on non-ferroelectric surfaces and leading to well-known difficulties in interpretation of the PFM data and hysteresis loops, as explored in depth over the last two decades.[25-28] However, we show that this explanation can be ruled out based on extensive set of measurements shown below, and a much more unusual and novel mechanism is realized in these systems.

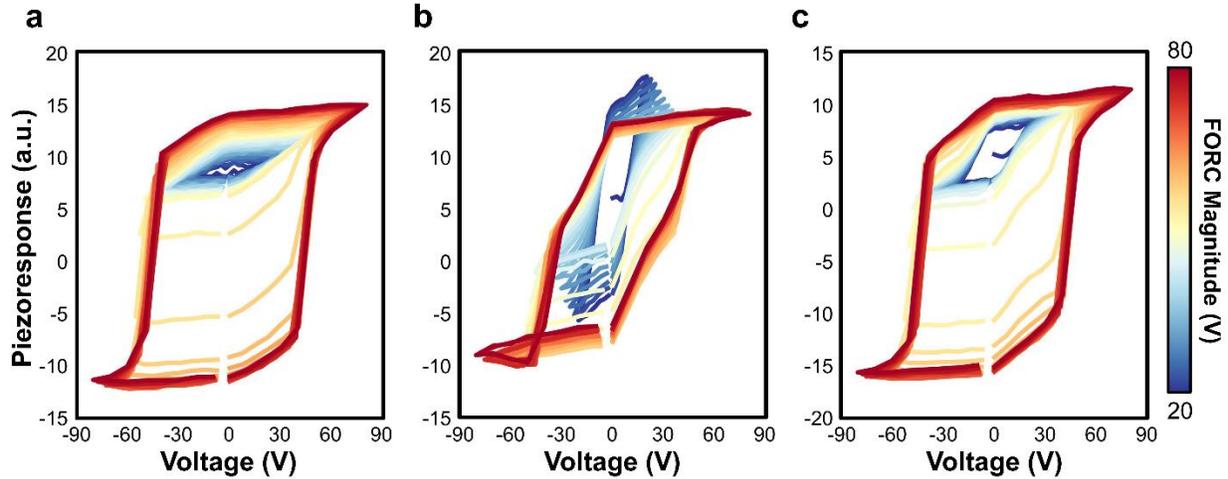

**Figure 2.** FORC hysteresis loops of three ZMO film. (a) ZMO-32. (b) ZMO-24. (c) ZMO-18.

As a next exploratory step, we probed polarization dynamics via switching spectroscopy measurements with acquiring of local hysteresis loops using BEPS. In these measurements, we applied a first order reversal curve (FORC) bias waveform, consisting of triangular pulses with gradually increasing amplitude, which allows to study the dynamics of polarization switching. The corresponding FORC waveform is shown in Figure S8 and comprises 20 cycles with amplitude ranging from 20 to 80 V. Figure 2 shows hysteresis loops measured in ZMO films of three different compositions. Similar to resulted domains, hysteresis loops show two stage switching process, resulting in presence of small loops (loop-1) with bias amplitude up to ~50 V and large loops (loop-2), where the small loop is located at the top side of the large loop in the pristine films. The small loop is stable when the DC voltage is smaller than 50 V, it quickly opens when the applied voltage is greater than 50 V and the loop opening is completed prior to the applied voltage reaching 60 V, as shown in Figure S9. Accompanied by loop opening, the coercive voltage also increases from ~10 to ~48 V (Figure S9).

Giving both domains and hysteresis loops show two states, we propose that domain-1 and domain-2 links to loop-1 and loop-2, respectively. To check this, we design an experiment that includes a DC set-pulse (SP) before FORC-BEPS measurement, as shown in Figure 3a. We modified the set-pulse in an automated manner using AEcroscoPy to perform a high throughput exploration of the set-pulse effect on FORC-BEPS hysteresis loops. The set-pulse array we used here is the same as the DC pulse array in our domain writing experiments, wherein the pulse magnitude ranges from negative 40 V-70 V and duration ranges from 0.1 ms to 5000 ms. High-throughput domain writing along with high-through set-pulse FORC-BEPS enable a direct comparison of the hysteresis loops and domains. The SP-FORC-BEPS results are shown in Figure 3b. It is observed in Figure 3b that the small hysteresis loops move from the top side to the bottom side within the large hysteresis loops when the set-pulse magnitude and duration increase. A few representative hysteresis loops that indicate the transition of the small loop location are amplified in Figure 3b. A comparison of the high throughput hysteresis loops with a set-pulse and the domain images reveals a correspondence between the small loops position and domain formation, that is, with the formation of a normal domain structure (the bottom right of the high throughput domain map in Figure 1b) the small loops move toward the bottom side of the large loops. These results unambiguously establish the connection between domains and hysteresis loops, the set pulse defines initial state in hysteresis loops and corresponding domains. Hereinafter, we refer the domain-1 and small hysteresis loop as ferroelectric-subsystem-1 (FE1), the domain-2 and large hysteresis loop as ferroelectric-subsystem-2 (FE2).

These studies also demonstrate that that the location change of the small loop is reversible, as shown in Figure 3c, the small loop can be moved back to the top side by applying a positive pulse, indicating the independence of the small loop. The independence of two sub polarization systems is very unusual, since the polarization distributions are strongly coupled through the depolarization field, and hence are expected to switch jointly. The high throughput SP-FORC-BEPS measurements are also performed on ZMO-24 and ZMO-18 films, which show reproducible observations, as shown in Figure S10.

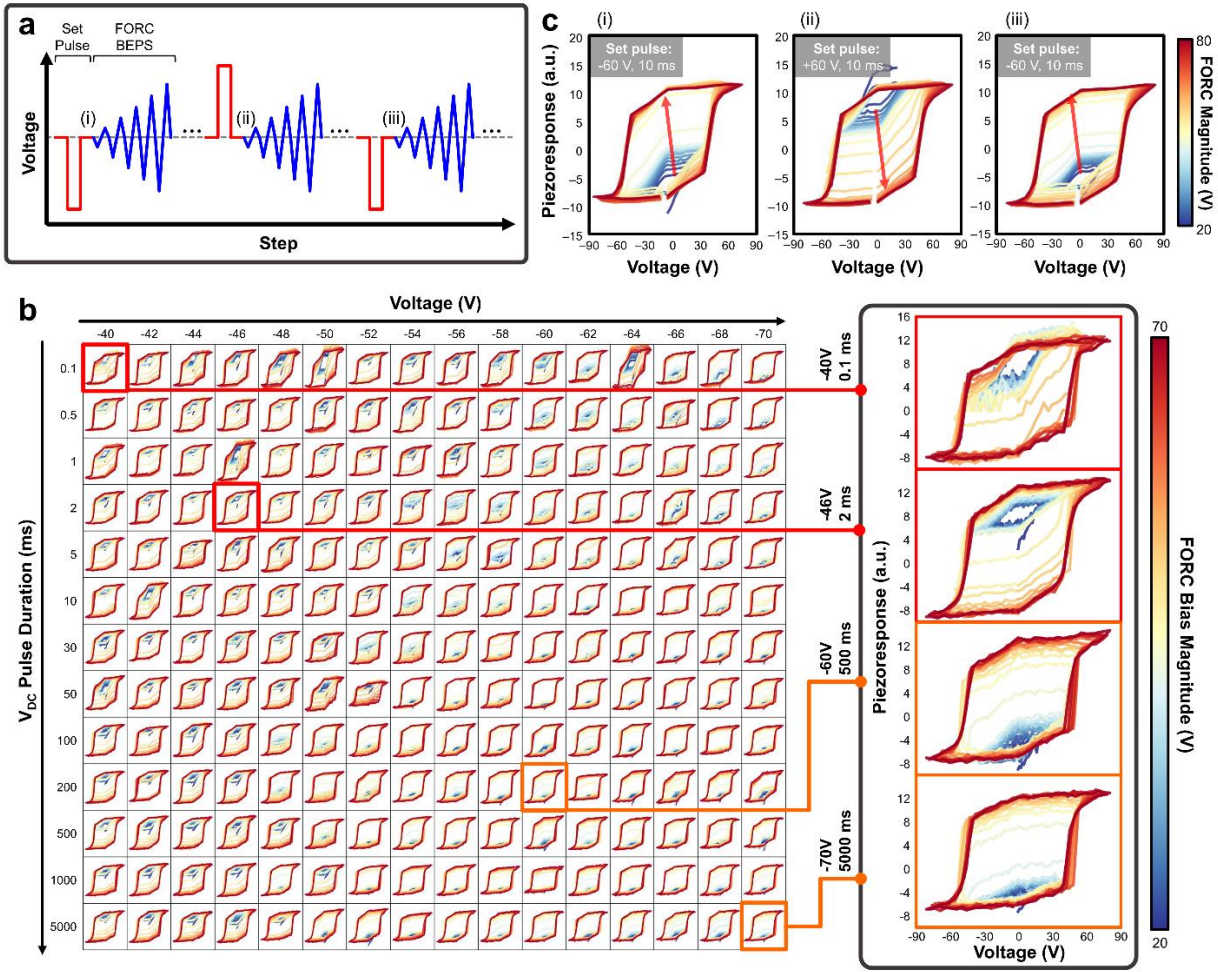

**Figure 3.** SP-FORC-BEPS hysteresis loops. (a), the measurement waveform, which comprises a set-pulse before FORC waveform. (b) high-throughput exploration of set pulse effects on hysteresis loops, a few representative loops are highlighted and amplified. (c) reversible change of the small loop location by applying positive and negative set-pulses.

We also note that while two-step hysteresis loops have been observed before, the past models cannot explain the observation in ZMO here. For example, complex PFM loops in $BiFeO_3$ and $H-LiNbO_3$ are attributed to defect dynamics, however defect dynamics cannot explain the movement of loop-1 location and the presence of domain-1. The surface charging can be an explanation of shadow domains, but not hollow domains. Another explanation, such as four-well potential previously reported in $CuInP_2P_6$ (CIPS),[29] is unlikely due to the simplicity of ZMO unit cell to form multiple well potential. Furthermore, it is important to interpret the observed phenomena in the context of unusual polarization dynamics in ZMO and similar binary oxide or nitride wurtzite systems.

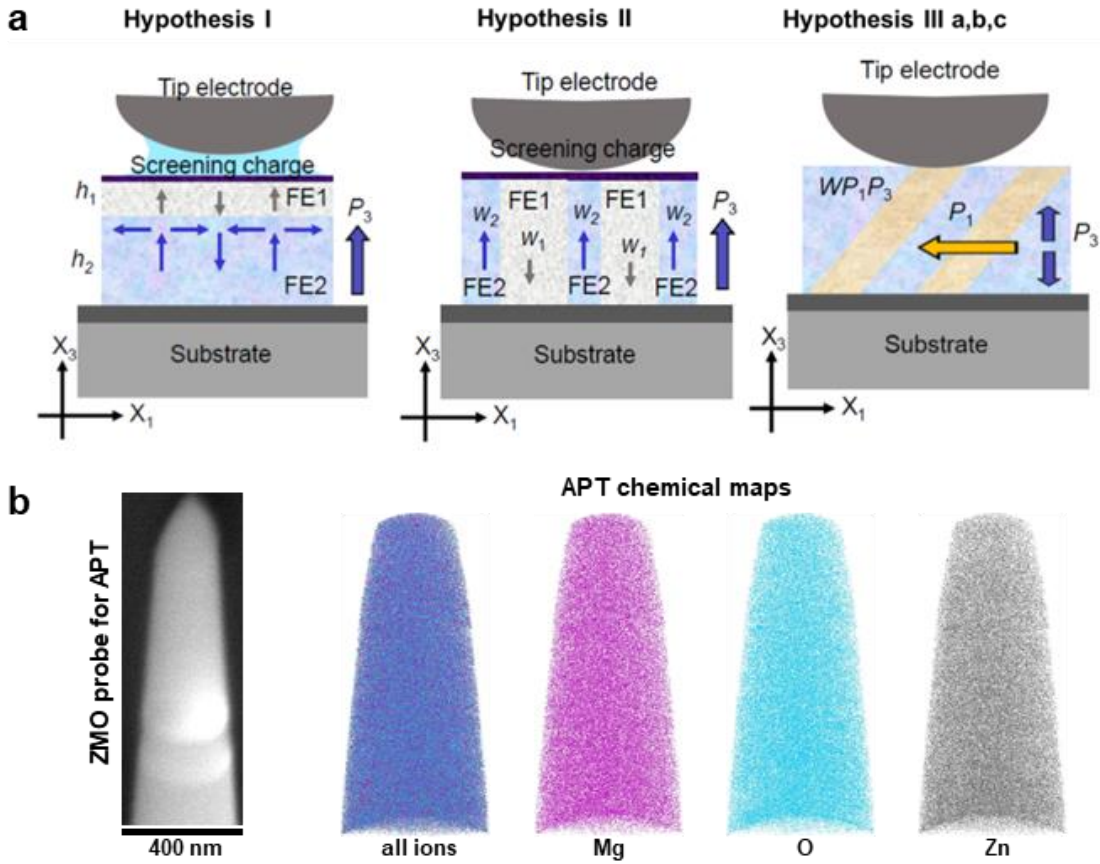

**Figure 4.** Hypotheses of two ferroelectric subsystems. (a) three possible hypotheses regarding two subsystems with different compositions, Hypothesis I: surface charging or selvedge properties, Hypothesis II: vertical tubular structures, Hypothesis III: presence of nano phase separation or two order parameters. (b) An SEM image of the prepared sample for the APT experiment, and the APT atom maps showing a homogenous elemental distribution.

To get insight into the observed phenomena, we analyze possible scenarios for polarization switching. We first consider that the two ferroelectric subsystems can have three possibilities based on geometric localization, as shown in Figure 4a. The hypotheses are as follows:
- hypothesis I: surface charging or selvedge properties,
- hypothesis II: vertical tubular structures with different compositions that switch independently,

- hypothesis III: bulk responses due to four well potential, presence of two order parameters, nano phase separation, or unusual domain dynamics.

To systematically explore these possibilities, we performed a set of depth- and spatially resolved characterization. As a first step, we performed Atom Probe Tomography (APT) to investigate whether there are tubular structures or phase separation. APT provides 3D compositional mapping with sub-nanometer resolution,[30, 31] enabling unique insights into the phase separation and composition variation. Shown in Figure 4b are the APT results that do not display any compositional variation, which was also confirmed using a frequency distribution analysis (FDA). This result is consistent with a homogeneous atomic distributions across the cross-section of the film as also observed in energy dispersive X-ray spectroscopy (EDX) elemental maps (Figure S11). Similarly time-of-flight secondary ion mass spectrometry didn't show any chemical variation through-out the bulk of the film (Figure S12). This rules out hypothesis II of vertical tubular structure and hypothesis III of bulk phase separation.

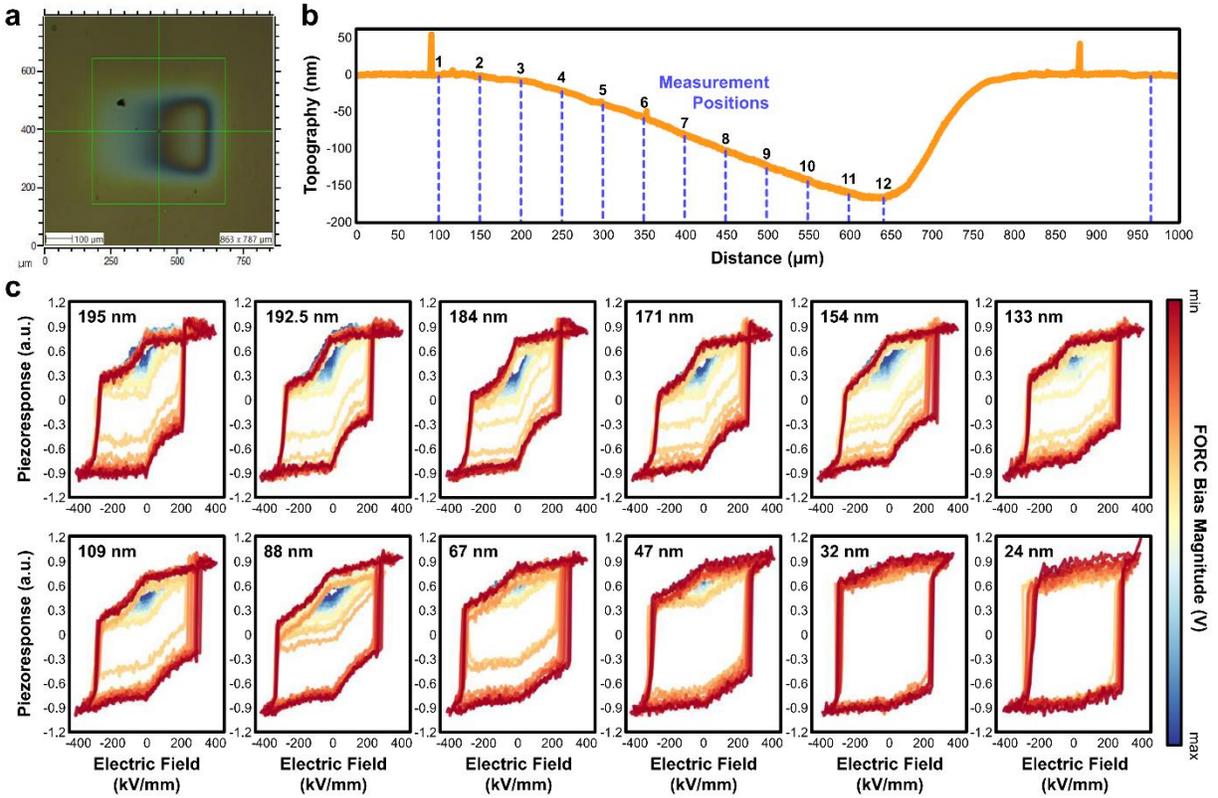

**Figure 5.** FORC hysteresis loops at a crater structure. (a) The crater structure is created in ToF-SIMS by gradually removing the surface layer of ZMO to form a thickness gradient. (b). AFM topography profile of the crater structure. (c). hysteresis loops at various remaining thickness, where the small loop disappears when the remaining thickness smaller than 32 nm.

To test the possibility of surface charging and surface electrochemistry, we used ion beam lithography to create a crater structure with a depth gradient by partially removing the materials, as schematically shown in Figure 5a. The AFM topography of the crater profile is displayed in Figure 5b. Shown in Figure 5c are hysteresis loops measured at different positions of the crater, each corresponds to a different depth. Noteworthily, ion beam lithography and PFM hysteresis loops are measured *in-situ* in an ultrahigh vacuum condition. If the double hysteresis loops are due to surface electrochemistry, it would disappear after the surface layer is removed. However, the double loops persist until the residual thickness is as low as 47 nm. The small loops disappear when the residual thickness is below 32 nm, accompanied with the disappearance of the kink in the large loop. This clearly evidences that the double loops are a bulk behavior, and cannot be attributed to surface electrochemical phenomena.

As a next step, we consider possible bulk models, including the presence of multiple order parameters and low symmetry. We were able to reproduce the experimental PFM behavior by introducing monoclinic materials (see Suppl Mat., part II.E). However, this model is falsifiable and predicts formation of complex domain patterns during switching that does not agree with the lateral PFM experiment as shown in Figure S13 (see the domain structures in Figs. S22-S23).

**Table 1.** Summary of hypotheses and corresponding evidence or refutation experimental results.

| Possible mechanism | Evidence/likelihood | Refutation experiment |
| --- | --- | --- |
| Hypothesis I: Surface electrochemical coupling and/or double layer subsystems | The PFM studies: strong and weak domains, hysteresis loops, independent switching of minor and major loops | FORC PFM spectroscopy in crater structure: thickness evolution of PFM spectra during subtractive modification |
| Hypothesis II: Vertical tubular structures | Two coexisting ferroelectric states require weak coupling via depolarization fields | APT and EDS indicate no chemical variation. PFM indicates no nano domains |
| Hypothesis III: Presence of two order parameters | Alternative explanation for the PFM studies | Lateral PFM does not show in-plane component, observed contract is crosstalk between vertical and lateral signals |
| Hypothesis IV: Observation of the pre-switching state | Remaining explanation | FORC PFM spectroscopy in crater structure: small loop disappears under a threshold thickness |

Consequently, as summarized in Table 1, this suggests that the remaining explanation is the fringing-ridge model, as illustrated in Figure 6. Switching in normal ferroelectrics starts with nucleation, vertical and lateral growth of domain, break-through to the bottom electrode, and lateral domain growth, as schematically shown in Figure 6a. In ZMO, the scenario is that the partial switching forming the fringing-ridge domains near the film surface. Before a critical bias, the fringing-ridge domain does not grow laterally and rarely propagates vertically. Rather, on increasing pulse magnitude and duration, more fringing-ridge domains nucleate next to the original one. This corresponds to our observation of domain-1 in the high-throughput domain writing PFM results in Figure 1, where observed domain-1 structure is not individual domain but rather many small needle-like domains, these needle-like domains show suppressed PFM amplitude because

PFM averages signal from regions of different polarities (needle-like domains and areas around them). After sufficiently high bias, one of the fringing-ridge domains, either a domain directly below the tip or a domain further away, breaks through to the bottom electrode. This corresponds to the observation of domain-2 in Figure 1. After breaking through the bottom electrode, the domain starts to grow laterally. The fringing-ridge mechanism is schematically shown in Figure 6b with representative PFM image corresponding to each step in the polarization reversal process.

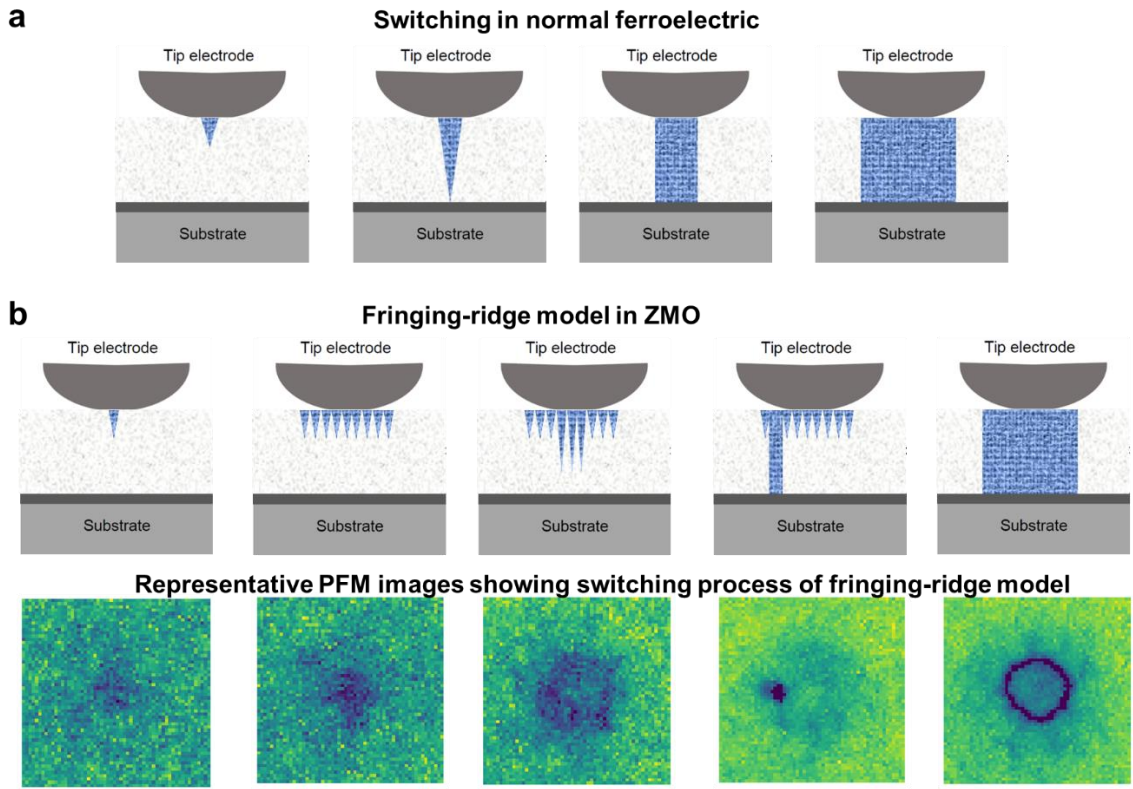

Figure 6. Schemes of polarization switching mechanisms. (a) polarization switching in normal ferroelectrics; (b) fringing-ridge mechanism in ZMO and representative PFM images corresponding to the switching process of fringing-ridge model.

Finally, we proceed to develop the semiqualitative model that can explain the observed dynamics. Following the logic behind the hypotheses 1-3, we consider in detail the polarization reversal in an inhomogeneous structure "ferroelectric FE1 - ferroelectric FE2" (see Suppl.Mat, part II), providing that the effective piezoresponse is proportional to the spontaneous polarization. Note that the verification of the possible bilayer-type, columnar-type or cross-type spatial orientations of FE1 and FE2 layers/regions can be performed experimentally. Since the spatial separation in chemical composition is absent, columnar-type and cross-type geometries of the FE1 and FE2 regions should be excluded, and so only bilayer-type geometries, shown in Figure S14a and b, remained for our consideration.

Below we consider the model when the concentration of free carriers and hence screening capability in layers "1" and "2" is different. Within our model the screening properties is defined by the spatial distribution of the Debye-Hückel screening radius. Due to diffusion, the bottom layer can be enriched by electrons coming from the bottom electrode, so that the electron concentration can be significantly higher in the 10-20 nm thick layer near the bottom electrode than outside this layer.

Since the exact distribution of these trap states densities is unknown, we consider several distributions of $R_d(x_3)$, namely the sharp step (solid curve), the diffuse step (dotted curve), and the maximum at the interface (dashed curve; see Figure 7b). We assume that a thin layer at the bottom, e.g. 20-nm, thick has a non-stoichiometric chemical composition with the maximal deviation from the stoichiometry in the central part of the layer (say at distance 10 nm from the bottom electrode).

To determine the spatial-temporal evolution of the polarization in the ZMO we use the Landau-Ginzburg-Devonshire (LGD) approach.[32] Corresponding LGD free energy functional $F$ additively includes a bulk part – an expansion on the 2-th, 4-th, 6-th and 8-th powers of the ferroelectric polarization component $P_3$ and its interaction with electric field, $F_{bulk}^{top}$ and $F_{bulk}^{bot}$; a polarization gradient energy contribution, $F_{grad}$, surface and interface energies, $F_{s+int}$. It has the form:

$$F = F_{bulk}^{top} + F_{bulk}^{bot} + F_{grad} + F_{s+int}. \quad (1)$$

In Eq.(1) the expansion coefficients ($\alpha, \beta, \gamma, \delta$) on the polarization powers are different for the top layer "1" free energy, $F_{bulk}^{top}$, where the mismatch and/or Vegard strains are absent, and for the bottom layer "2" free energy, $F_{bulk}^{bot}$, where the strong and anisotropic strains can be induced by oxygen vacancies and ions migration, and/or by the substrate. The parameters of the top layer correspond to the spontaneous polarization and coercive field of a bulk ZMO.[7, 33] The surface energy via the random variations of its coefficients can define the features of the domain nucleation, and, in particular, the appearance of the "fringing ridge". The spatial-temporal evolution of the ferroelectric polarization is determined from the coupled time-dependent LGD type Euler-Lagrange equations, obtained by the minimization of the free energy $F$, $\Gamma_P \frac{\partial P_3}{\partial t} = -\frac{\partial F}{\partial P_3}$.

In this case, the Euler-Lagrange equations with different LGD coefficients for polarization are valid for both layers; and electric fields in the layers differ due the influence of charge carriers considered in the Debye-Hückel approximation. Dynamic hysteresis loops of local polarization $P_3$ calculated for a slowly increasing amplitude of the tip bias (from red to magenta curves), increasing thickness $h_1$ from 10 nm to 60 nm and fixed $h_2$=20 nm is shown in Figure 7c.

It is seen from Figure 7c, that the increase of the top layer thickness within the range 10 nm $\leq h_1 \leq$ 30 nm leads to the opening of a "big" major loop with a high coercive bias, and to the simultaneous appearance of two "small" minor loops inside the major loop. Note that such minor loops seem symmetry-protected in uniaxial ferroelectrics.[34] The minor loops are observed when the top and bottom layers have close thicknesses, $h_1 \sim h_2$. Furter increase of the top layer thickness within the range $h_1 \geq$ 40 nm leads to the disappearance of the minor loops, and to the simultaneous increase of the coercive bias corresponding to the major loop. The presence of random surface potential induces visible irregularity of the domain structure and shifts the local polarization loops in both vertical and lateral directions. It seems more important that the surface potential breaks the symmetry of the two minor loops making one of them bigger and more pronounced. Black points with numbers in Figure 7d denote the stages of polarization reversal in the bilayer structure with $h_1 = h_2$ =20 nm, for which the cross-sections of the polarization distributions are shown. It is seen that the fringing ridge can originate, grow and intergrowth through both layers. It is important to note that the PFM signal is sensitive to the molar fraction of the material with a given polarization orientation averaged over the PFM signal generation volume, the probing depth in PFM is comparable to ½ of contact radius. As such, the double layers can consist a surface layer and a bulk layer that is detectable by PFM; in this manner, two ferroelectric subsystems can in principle originate from surface and bulk, respectively.

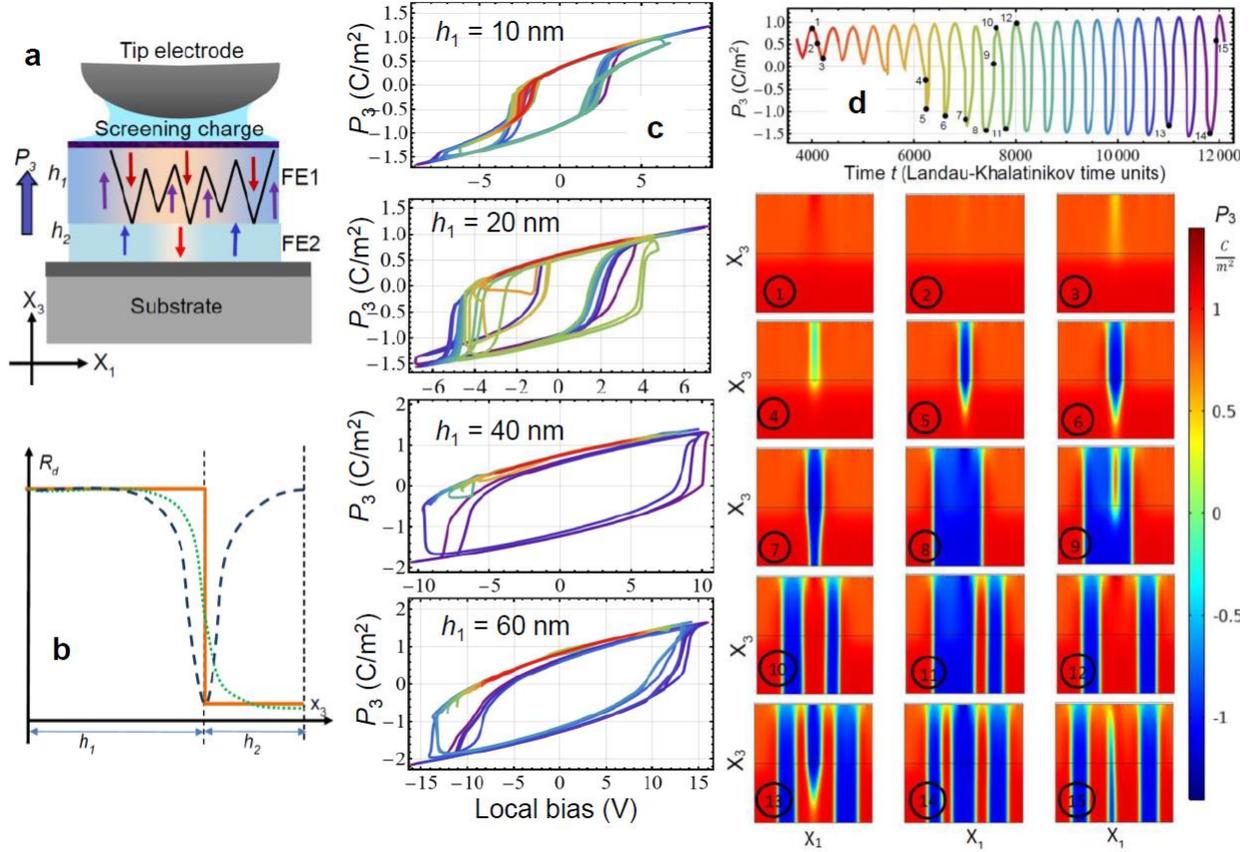

**Figure 7.** (a) Geometry of the considered models, "ferroelectric FE1 - ferroelectric FE2", describing an inhomogeneous ZMO film between the conducting tip and the bottom electrode/substrate. (b) Several hypotheses (solid, dashed and dotted curves) for $R_d(x_3)$ distribution. (c) Dynamic hysteresis loops of local polarization $P_3$ calculated for a slowly increasing amplitude of the tip bias (from red to magenta curves) and increasing thickness $h_1$ of ZMO bilayer, $h_2=20$ nm. (d) Polarization distributions below the probe tip calculated in the cross-section of the are calculated for different stages of polarization reversal in the points denoted by numbers 1-15 at the time dependence of local polarization; $h_1 = h_2 = 20$ nm.

This analysis illustrates that the correlated spike-like domains, named as "fringing-ridge", can appear during the polarization reversal in the ZMO films. They are induced by a chemical pressure (Vegard strains), or/and other types of elastic defects, which concentration increases significantly near the surfaces and interfaces. The fringing-ridge, like shown in Figure 7a, can act as the new order parameter,[32] which can transform the two-well potential of ZMO into the multi-well potential. The amount and the average width of the teeth are defined in a self-consistent way,[32] because these quantities determine the structure of the "effective" multi-well potential. As shown previously,[34] a specific evolution of the potential wells can lead to the changes in the hysteresis loops topology from a single loop to the complex loops, similar to the ones observed in ZMO.

To summarize, we report highly unusual ferroelectric behavior of $Zn_{1-x}Mg_xO$, the only self-consistent model aligned with all experimental evidence is one where two independent ferroelectric subsystems coexist in an initially layered geometry. Using band excitation piezoresponse force microscopy (BEPFM) and spectroscopy (BEPS), the research highlights distinct polarization switching dynamics, contrary to traditional depolarization field coupling theories. The discovery of a fringing-ridge mechanism, characterized by lateral nucleation expansion without significant depth propagation, presents a novel type of ferroelectric switching. This diverges from the usual domain growth process observed in ferroelectrics. This could be indicative of more complex dynamics in similar binary oxide or nitride wurtzite systems.

The implications of these findings are significant for the semiconductor and information technology industries, both by linking to the macroscopic switching dynamics and suggesting strategies for materials optimization. The unique polarization behavior of $Zn_{1-x}Mg_xO$, especially the fringing-ridge mechanism, offers insights into designing more efficient and scalable ferroelectric materials. These materials are vital for applications like high-density information storage and ferroelectric random-access memories (FeRAMs). The study enhances understanding of ferroelectric materials' compatibility with CMOS technology, potentially paving the way for innovative memory solutions with low power dissipation, fast operational speeds, enhanced endurance, and improved retention.

**Methods**

*Materials Synthesis*

ZMO samples are fabricated on Pt bottom electrodes on sapphire substrates. Pt is deposited by RF magnetron sputtering at 350 °C using a 5 nm Ti adhesion layer. ZMO films are reactively RF magnetron sputtered from 2" diameter Zn and Mg metal targets at room temperature. Both targets are 99.99% pure on a metals basis (Kurt J. Lesker). Gun power settings establish the Zn:Mg ratio. The sputtering atmosphere is a 1 mTorr mixture of 6:1 Ar:$O_2$. The magnetrons are oppositely opposed with a 45° angle from the substrate normal, target-to-substrate distance is approximately 5 cm. Deposition rates vary for different Zn:Mg ratios but are in the range of 15 nm/min. 4-circle x-ray diffraction (Panalytical Emperyan) is used to quantify film crystallinity, phase, orientation and thickness. Zn:Mg ratios are measured using energy dispersive x-ray spectroscopy (Oxford Instruments).

*BEPFM and BEPS*

BEPFM and BEPS were measured on a commercial Cypher AFM system (Asylum Research, Oxford Instruments Co.), which is equipped with a band-excitation generator. A Pt/Ir coated AFM tip (ElectriMulti75-G (Budget Sensors)) with a nominal stiffness of $3\,\text{N}\,\text{m}^{-1}$ was used. The automated experimentation was achieved via AEcroscoPy[19] reported recently, where the experimentation workflow is programmed with Python and execution was realized via Python-LabView-Microscope integration.

*ToF-SIMS and PFM*

ToF-SIMS measurements were carried out using TOF.SIMS.5-NSC instrument combining AFM and ToF-SIMS in the same vacuum chamber. Bismuth liquid metal ion gun (30 keV energy, 0.5 nA current in DC mode and ~120 nm spot size) was used as a primary ion source for extracting of the sample secondary ions. Cs sputter source (1 keV energy, ~80 nA current) was used for depth profiling and oxygen sputter source (500 eV energy, ~100 nA current) was used for wage sputter. Time-of-flight mass analyzer was used for detection of the secondary ions in negative ion detection mode and allowed mass resolution m/Δm = 100 - 300. PFM and switching spectroscopy measurements on the crater structure were carried out using AFM in the same vacuum chamber with ToF-SIMS, using a suit of external electronics.

*APT*

Standard lift-out and annular FIB milling procedures were done using an FEI Nova 200 Dual-Beam FIB-SEM instrument to prepare the APT specimens.[35] The APT experiments were conducted using a CAMECA LEAP 4000XR in laser mode with a 30K base temperature, a 50 pJ laser energy, a 125 kHz pulse repetition rate, and a 0.5% detection rate. The reconstruction and data processing were performed using CAMECA's IVAS 3.8 software package.

**Acknowledgements**

This effort (materials synthesis, PFM measurements) was supported as part of the center for 3D Ferroelectric Microelectronics (3DFeM), an Energy Frontier Research Center funded by the U.S. Department of Energy (DOE), Office of Science, Basic Energy Sciences under Award Number DE-SC0021118. The PFM, APT, and ToF-SIMS were supported by the Center for Nanophase Materials Sciences (CNMS), which is a US Department of Energy, Office of Science User Facility at Oak Ridge National Laboratory. ToF-SIMS characterization was conducted at the Center for Nanophase Materials Sciences, which is a DOE Office of Science User Facility, and using instrumentation within ORNL's Materials Characterization Core provided by UT-Battelle, LLC under Contract No. DE-AC05-00OR22725 with the U.S. Department of Energy. Sergei V. Kalinin acknowledges support from the Center for Nanophase Materials Sciences (CNMS) user facility which is a U.S. Department of Energy Office of Science User Facility, project no. CNMS2023-A-01923. A.N.M. work is supported by the Ministry of Science and Education of Ukraine (grant № PH/ 23 - 2023, "Influence of size effects on the electrophysical properties of graphene-ferroelectric nanostructures") at the expense of the external aid instrument of the European Union for the fulfillment of Ukraine's obligations in the Framework Program of the European Union for scientific research and innovation "Horizon 2020". The authors would like to thank James Burns for assistance in performing APT sample preparation and running the APT experiments. The authors would like to acknowledge helpful discussion with Sabine M. Neumayer.

**Authors Contribution**

J.Y., S.V.K, and Y.L. conceived the project and co-wrote the manuscript. Y.L. developed PFM experimental methods. J.Y. and Y.L. performed ambient PFM measurements and results analysis. A.V.I. created the crater structure and performed PFM and ToF-SIMS in UHV. D.G, R.J.S, and J-P.M synthesized materials. J.D.P performed APT characterization. S.V.K, Y.L., A.V.I, and A.M.N. developed hypotheses and A.M.N and E.A.E. performed simulation. All authors edited the manuscript.

**Conflict of Interest**

The authors declare no conflict of interest.